\documentclass[twocolumn]{aastex62}

\usepackage{xcolor}

\received{xxx xx, 2018}
\revised{xxx xx, 2018}
\accepted{xxx xx, 2018}
\submitjournal{ApJ Letters}

\shorttitle{Did ASAS-SN kill PG1302-102?}
\shortauthors{T. Liu et al.}

\begin{document}

\title{Did ASAS-SN Kill the Supermassive Black Hole Binary Candidate PG1302-102?}

\author{Tingting Liu}
\affiliation{Department of Astronomy, University of Maryland, College Park, Maryland 20742, USA}
\affiliation{tingting@astro.umd.edu}

\author{Suvi Gezari}
\affiliation{Department of Astronomy, University of Maryland, College Park, Maryland 20742, USA}

\author{M. Coleman Miller}
\affiliation{Department of Astronomy, University of Maryland, College Park, Maryland 20742, USA}

\begin{abstract}
\cite{Graham2015Nature} reported a periodically varying quasar and supermassive black hole binary candidate, PG1302-102 (hereafter PG1302), which was discovered in the Catalina Real-Time Transient Survey (CRTS). Its combined Lincoln Near-Earth Asteroid Research (LINEAR) and CRTS optical light curve is well fitted to a sinusoid of an observed period of $\approx 1,884$ days and well modeled by the relativistic Doppler boosting of the secondary mini-disk \citep{D'Orazio2015}. However, the LINEAR+CRTS light curve from MJD $\approx 52700$ to MJD $\approx 56400$ covers only $\sim 2$ cycles of periodic variation, which is a short baseline that can be highly susceptible to normal, stochastic quasar variability \citep{Vaughan2016}. In this Letter, we present a re-analysis of PG1302, using the latest light curve from the All-Sky Automated Survey for Supernovae (ASAS-SN), which extends the observational baseline to the present day (MJD $\approx 58200$), and adopting a maximum likelihood method which searches for a periodic component in addition to stochastic quasar variability. When the ASAS-SN data are combined with the previous LINEAR+CRTS data, the evidence for periodicity decreases.  For genuine periodicity one would expect that additional data would strengthen the evidence, so the decrease in significance may be an indication that the binary model is disfavored.
\end{abstract}



\section{Introduction} \label{sec:intro}

Periodic light curve variability of quasars has been predicted as an observational signature of supermassive black hole binaries (SMBHBs) at sub-parsec separations, due to modulated mass accretion onto the binary (e.g. \citealt{D'Orazio2013, Gold2014, Farris2014}), or relativistic Doppler boosting of the emission of the secondary black hole mini-disk \citep{D'Orazio2015}. This predicted signature has motivated several systematic searches for periodically varying quasars in large time domain surveys, including \cite{Graham2015Nature} (hereafter G15), \cite{Graham2015}, \cite{Liu2015}, \cite{Liu2016}, and \cite{Charisi2016}, and spurred a number of recent claims of (quasi-)periodicity (and binarity) that were discovered serendipitously or in previously well-known AGN\footnote{However, some of these claims have already been challenged: for example, Barth \& Stern (2018) pointed out some issues that affect the Dorn-Wallenstein et al. (2017) analysis.} (e.g. \citealt{Dorn-Wallenstein2017,Kova2018}). G15 reported a periodic quasar and SMBHB candidate PG1302-102 (hereafter PG1302), whose light curve from the Catalina Real-Time Transient Survey (CRTS) can be fitted to a sinusoid of an observed period of $P = 1884\pm88$ days over the $\sim 9$-year CRTS baseline. Its light curve including the Lincoln Near-Earth Asteroid Research (LINEAR; \citealt{Sesar2011}) data, which extends $\sim 0.5$~cycles before the CRTS data, is consistent with the sinusoidal fit, and archival photometry data from various telescopes are largely consistent with the extrapolation of the sinusoid $\sim 10$ years before LINEAR, although their sampling is sporadic.

While there have been multi-wavelength analyses of PG1302 in the UV \citep{D'Orazio2015}, IR \citep{Jun2015}, and radio \citep{Kun2015}, which can provide key complementary clues about the true nature of a variability-selected SMBHB candidate, the periodicity of PG1302 remains unconvincing due to the small number of cycles (N$_{\rm cycle}$ $\sim 2$ over a combined LINEAR+CRTS baseline). \cite{Vaughan2016} have cautioned against claiming periodicity over such a small number of cycles, as the stochastic variability (``red noise'') of normal quasars and AGN (i.e., those that do not host SMBHBs) can easily mimic periodic variation. Indeed, \cite{Vaughan2016} showed that aperiodic light curves simulated using the Damped Random Walk model (DRW; \citealt{Kelly2009}) or a broken power law (BPL) power spectrum can also be fitted to few-cycle data after down-sampling and adding photometric noise. Moreover, an extended baseline analysis using new monitoring data disfavors the persistence of the periodic quasar candidates from the Pan-STARRS1 Medium Deep Survey (PS1 MDS) MD09 field \citep{Liu2016}.

Three years after G15 and five since its last published CRTS data, we revisit the periodicity of PG1302 in this Letter, by adding the publicly available light curve from the All-Sky Automated Survey for Supernovae (ASAS-SN). We describe the ASAS-SN light curve in Section \ref{sec:asas} and the maximum likelihood method we use in the analysis in Section \ref{sec:zoghbi}. In Section \ref{sec:ext} we describe and simulate the expectations in the case where a genuine periodicity is present, and then compare those expectations with our reanalysis of PG1302. We conclude in Section \ref{sec:conclude}.


\section{Extended light curve from ASAS-SN}\label{sec:asas}

The ASAS-SN survey \citep{Shappee2014,Kochanek2017} is regularly monitoring the variable sky down to V $\sim 17$ mag using multiple telescopes hosted by the Las Cumbres Observatory. We retrieved the ASAS-SN light curve of PG1302 (J2000 RA = $196.3875$, Dec = $-10.5553$) from 2012 February 15 to 2018 March 1 (MJD $= 55972 - 58178$) from the Sky Patrol\footnote{https://asas-sn.osu.edu}. For calibration purposes, we choose the length of the ASAS-SN light curve ($\approx 2,200$ days) to overlap with the CRTS light curve by $\sim 400$ days. Due to the dense sampling and the large photometric uncertainty of the ASAS-SN light curve, we have binned the light curve using a width of $\sim 100$ days (such that there are $20$ bins over $\sim 2000$ days with an average of $46$ measurements per bin) using the arithmetic mean, and the uncertainty of each bin is given by the standard deviation of the measurements.

The CRTS \citep{Drake2009} light curve of PG1302 was retrieved from the Second Data Release of the Catalina Sky Survey (CSS). While V$_{\rm CSS}$ is based largely on the Johnson V magnitude system used in ASAS-SN, there are some differences. Instead of calculating a color-dependent correction to convert between the V magnitudes of the two surveys, we simply apply a constant offset to the ASAS-SN light curve before it was ``stitched'' to the CRTS light curve: after binning the CRTS data via the same method described above ($15$ bins each of width of $\sim 180$ days), we calculate the difference between the (binned) CRTS and ASAS-SN magnitudes in each of the two overlapping seasons, i.e., MJD $\approx 55900 - 56100$ and MJD $\approx 56200 - 56500$, and offset the ASAS-SN light curve by the average difference ($0.17$ mag) in order to match to CRTS. The LINEAR light curve of PG1302 has also been offset and binned in the same way.  Although early-time data from \cite{Garcia1999}, \cite{Eggers2000} and ASAS \citep{Pojmanski1997} are generally consistent with the extrapolated sinusoidal fit to LINEAR+CRTS data, we do not include them in our analysis due to their much sparser sampling and less reliable photometry.

\begin{figure*}[h]
\plotone{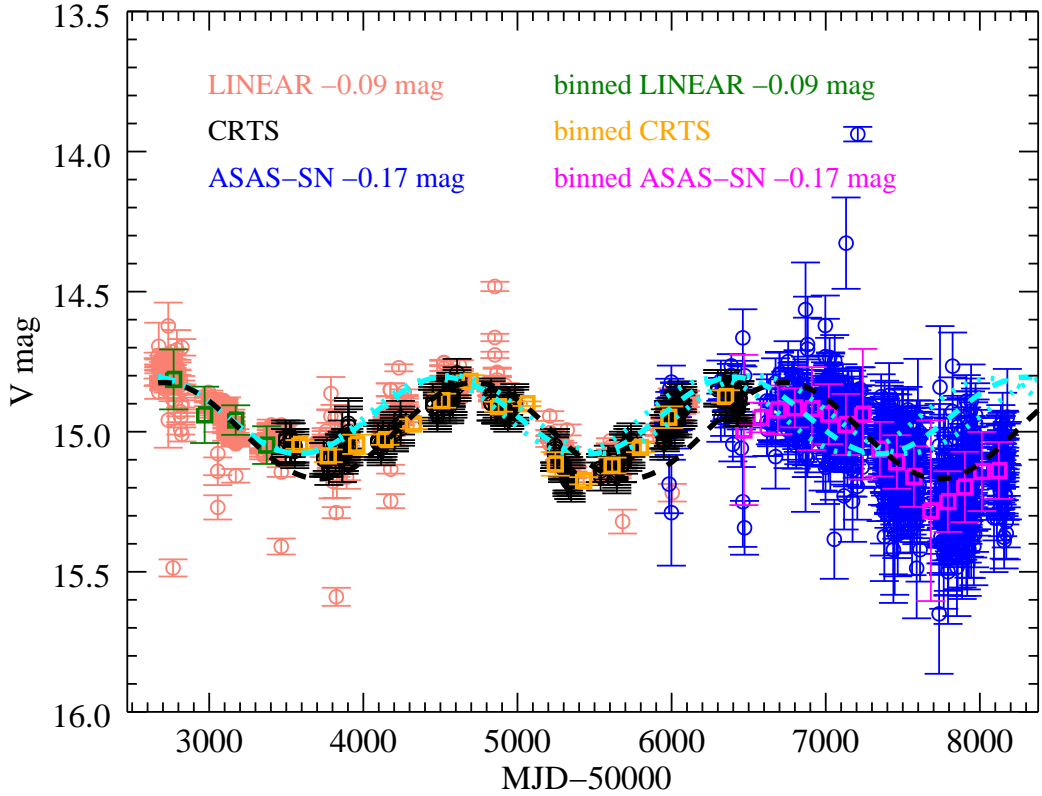}
\caption{The combined light curve of PG1302-102 from LINEAR (pink), CRTS (black) and ASAS-SN (blue). LINEAR and ASAS-SN have been offset to match CRTS (see text). Adopting the best-fit period and its uncertainties from G15, sinusoids with periods of P $= 1884$ days (cyan dashed line) and P $=1884\pm88$ days (cyan dotted lines) have been fitted to the LINEAR+CRTS light curve and extrapolated to guide the eye. Additionally, we have superimposed a best-fit sinusoid of the period P$=2012$ days (black dashed line), the best-fit period of the LINEAR+CRTS+ASASSN light curve that we determined under the DRW+periodic model. The binned light curve is also shown (LINEAR: green; CRTS: orange; ASAS-SN: magenta).}
\label{fig:ext_PG1302}
\end{figure*}

The full baseline in our analysis is therefore given by LINEAR+CRTS+ASASSN. We present both the binned and un-binned light curves in Figure \ref{fig:ext_PG1302}. Although the ASAS-SN light curve does undulate, the periodic fluctuation detected in the CRTS light curve is not consistent with the ASAS-SN data.  In particular, the extended ASAS-SN light curve fluctuation is clearly out of phase with the sinusoid fitted to the LINEAR+CRTS light curve, and the full data set favors a longer apparent period and larger amplitude.

\section{Expectations for a true periodic signal}\label{sec:zoghbi}

Since the LINEAR+CRTS+ASASSN light curve is inconsistent with a sinusoid of the best-fit period and phase from G15, we now analyze the combined data by considering a possible periodic signal in the presence of red noise. The basic picture is that fluctuations in the accretion disk can produce a red noise component in the power spectrum, whereas the binary is expected to produce a periodic signal.

We adopt the maximum likelihood method introduced by \cite{Bond1998}, which has been applied in a number of previous studies, including \cite{Miller2010}, \cite{Zoghbi2013}, and \cite{Foreman-Mackey2017}. The observed light curve is the combination of signal and noise: \textbf{x} = \textbf{s} + \textbf{n}, or in terms of a correlation matrix:

\begin{eqnarray}
C_{x} = C_{s} + C_{n}\quad,
\end{eqnarray}

\noindent where $C_{s} = \langle s_{i}s_{j}\rangle$ and $C_{n} = \langle n_{i}n_{j}\rangle$, and the indices $i$ and $j$ indicate elements of the light curve, which has a total of $N$ elements. The noise terms are assumed to be Gaussian (which is usually true in optical astronomy); further assuming that they are uncorrelated, $C_{n}$ is simply a diagonal matrix with elements $n_{i}n_{i}$. Each element of the signal matrix $C_{s}$ can be expressed using the autocorrelation function:

\begin{eqnarray}
\langle s_{i}s_{j}\rangle = \mathcal{A}(\Delta t) = \int_{-\infty}^{+\infty} P(f) \cos(2\pi f\Delta t) df\quad,
\label{eqn:a}
\end{eqnarray}

\noindent where $P(f)$ is the power spectral density (PSD) of the signal, and $\Delta t$ is the time lag between $s_{i}$ and $s_{j}$. Having calculated the signal matrix $C_{x}$ for a set of parameters \textbf{p}, we can then construct a likelihood function $\mathcal{L} (\textbf{p})$ under the model $P(f)$:

\begin{eqnarray}
\mathcal{L} (\textbf{p}) = (2\pi)^{-N/2} |C_{x}|^{-1/2} \rm exp(-\frac{1}{2} \textbf{x}^{T} C_{x}^{-1} \textbf{x})\quad,
\label{eqn:lkhd}
\end{eqnarray}

\noindent where $|C_{x}|$ and $C_{x}^{-1}$ are the determinant and inverse of the matrix $C_{x}$, respectively, and \textbf{x}$^{T}$ is the transpose of the time series \textbf{x}. To calculate the likelihood under the Damped Random Walk model (DRW; \citealt{Kelly2009}), which has been successful in characterizing quasar variability (e.g. \citealt{Kelly2009,MacLeod2010}), $P(f)$ in Equation \ref{eqn:a} would take the following form: 

\begin{eqnarray}
P(f) = \frac{2\sigma^{2} \tau{^2}}{1+(2\pi \tau f)^{2}}\quad,
\label{eqn:drw}
\end{eqnarray}

\noindent where $\sigma^{2}$ is the short-timescale variance, and $\tau$ is the characteristic timescale. To search for a periodic component of frequency $f_{0}$ in addition to DRW noise (hereafter ``DRW+periodic''), we can introduce a delta function $\delta(f - f_{0})$, so that the autocorrelation function in Equation \ref{eqn:a} becomes:

\begin{eqnarray}
\mathcal{A}(\Delta t) = \Big[\int_{-\infty}^{+\infty} P(f) \cos(2\pi f\Delta t) df \Big]+ A_{0} \cos(2\pi f_{0}\Delta t)\quad.
\label{eqn:drw_p}
\end{eqnarray}

\noindent where $A_{0}$ is the amplitude of the periodic signal.

\begin{figure}[h]
\plotone{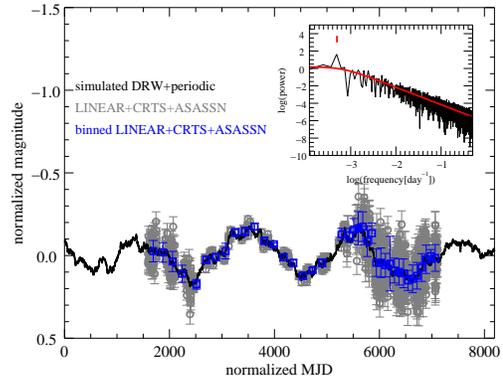}
\caption{We generate a light curve under the DRW model and inject a periodic function. The light curve is initially nightly sampled (black line). We then down-sampled the perfect light curve and added typical photometric noise of the LINEAR, CRTS, and ASAS-SN data (grey circles with error bars). The resampled light curve is then binned (blue squares with error bars). The inset shows the periodogram of the evenly-sampled light curve without photometric noise. The DRW model which generates the light curve is superimposed (red line), and the input period is indicated with a red tick mark. We find that despite the significant photometric uncertainties in the simulated ASAS-SN data, its addition to the analysis strongly improves the evidence for periodicity when a periodic signal is actually present.}
\label{fig:test_drw+p}
\end{figure}

To test our implementation of the method, we simulated ten light curves under the DRW model using the \cite{Timmer1995} method, uniformly sampling $\sigma$ from $0.00224$ mag day$^{-1/2}$, which is the minimum value from the \cite{Kelly2009} quasar sample, to $0.0187$ mag day$^{-1/2}$, which corresponds to the value at $3 \sigma_{\rm Gaussian}$ after fitting the \cite{Kelly2009} $\sigma_{\rm DRW}$ distribution to a Gaussian; the input $\tau$ ranges from $\approx 30-970$ days\footnote{All temporal parameters explored in this analysis are in the observed frame.}. We then add sinusoidal functions with amplitudes measured from the periodic candidates from PS1 MDS (Liu et al. in prep.) so that A$_{0} \approx 0.1 - 0.3$ mag. The input periods range from $P \approx 50-970$ days; the maximum period corresponds to $2/3$ of the length of the baseline, which is the requirement in previous work including \cite{Graham2015} and \cite{Charisi2016}. We then down-sample the light curve to the observing cadence of PS1 MDS and add typical PS1 photometric noise. We then use a \texttt{C} implementation of an affine-invariant MCMC sampler \citep{affine} to sample the parameter space. Our implementation is successful in recovering the input period: the best-fit periods generally follow a one-to-one correlation with the input values. To select those by which the DRW+periodic model is at least moderately preferred, we further impose the cut AIC$_{\rm DRW+periodic}$$-$AIC$_{\rm DRW}$ $<-2$, where the Akaike information criterion AIC $ = 2n-2\ln\mathcal{L}$ when there are $n$ parameters in the model.  The AIC imposes a penalty on the more complex model, and between two models the model with the lower AIC value is therefore the preferred one. Those best-fit periods that meet this criterion correspond to $>3$ cycles, and they follow a yet tighter correlation.

\begin{deluxetable*}{lrrrr}
\tablecaption{Maximum likelihoods for the simulated DRW+periodic light curve \label{tab:drw_sim}}
\tablehead{
\colhead{} & \colhead{DRW} & \colhead{DRW+periodic} & \colhead{DRW} & \colhead{DRW+periodic} \\
\colhead{} & \colhead{\tiny{(LINEAR+CRTS)}} & \colhead{\tiny{(LINEAR+CRTS)}} & \colhead{\tiny{(LINEAR+CRTS+ASASSN)}} & \colhead{\tiny{(LINEAR+CRTS+ASASSN)}}
}
\startdata
$\ln \mathcal{L}_{\rm max}$ & 22.98 & 30.04 & 46.42 & 58.17 \\
p-value & \nodata & 8.59$\times10^{-4}$ & \nodata & 7.87$\times10^{-6}$ \\
P$_{\rm best fit}$ (day) & \nodata & 2060.75$^{+229.75}_{-430.24}$ & \nodata & 2026.83$^{+59.42}_{-70.57}$ \\
\enddata
\end{deluxetable*}

Next, we apply the method to a simulated DRW +periodic light curve to demonstrate the expected decrease in the p-value (and therefore increase in significance) if the periodic signal is real. We down-sampled the simulated light curve to the sampling of the LINEAR+CRTS+ASAS-SN light curve and added photometric uncertainties that are typical of the three different surveys (Figure \ref{fig:test_drw+p}). The light curve is then binned using the same bin sizes as Figure \ref{fig:ext_PG1302}. The relative amplitudes of the sinusoid and DRW noise are such that the significance level at which the DRW+periodic model is preferred is comparable between the (binned) LINEAR+CRTS-sampled light curve from the simulation and that from PG1302. The input period of P = $2012$ days is chosen to be the same as the best-fit period from our reanalysis of the LINEAR+CRTS+ASASSN light curve of PG1302 (Section  \ref{sec:ext}), and the phase of the simulated light curve also mimics that of PG1302. As Table \ref{tab:drw_sim} shows,  the method consistently recovered the input period in the LINEAR+CRTS and LINEAR+CRTS+ASASSN-sampled light curves, and the longer baseline produced a best-fit period that is closer to the true value with a smaller uncertainty. Furthermore, the p-value (for a chi-squared distribution with two degrees of freedom) has decreased significantly (by a factor of $\sim 100$) when the mock ASAS-SN data are included, even though they have a larger photometric uncertainty than the simulated CRTS data.


\section{Extended baseline analyses of PG1302}\label{sec:ext}

We now apply this method to PG1302, and the ranges of the sampled parameters are summarized in Table \ref{tab:drw}; in particular, the ranges of $\tau$ and $P$ are sampled from $200$ days to $3000$ days (recall that the putative period is $P=1884$ days). Since calculating the inverse and determinant of a large $N\times N$ matrix is computationally intensive (Equation \ref{eqn:lkhd}; both are typically $\mathcal{O}(N^{3})$ operations\footnote{However, we note that the algorithm \texttt{celerite} \citep{Foreman-Mackey2017} is able to compute Equation \ref{eqn:lkhd} at a cost of $\mathcal{O}(N)$ for some classes of PSD models, which include DRWs.}), where $N\sim 1000$ for the unbinned full-baseline light curve, we apply the method only to the binned light curve, where $N=19$ for LINEAR+CRTS and $N = 35$ for LINEAR+CRTS+ASASSN.  When we first applied the method to the CRTS-only and LINEAR+CRTS light curves (Table \ref{tab:drw}), the DRW+periodic model is preferred over the DRW-only model at the $98.4$\% and $99.9$\% levels, respectively. If PG1302 were the only quasar analyzed, this would be intriguing evidence for periodicity. However, given that it was selected from an initial sample of $\sim 200,000$ CRTS quasars, its periodicity can easily be produced by chance alone; to demonstrate strong evidence for periodicity, the candidate should instead have a p-value $< 5\times10^{-6}$.

As we showed in Section \ref{sec:zoghbi}, for a genuinely periodic source we expect that additional data should strengthen the evidence. However, the p-value of the DRW+periodic model has increased from $p=1.39\times10^{-3}$ on the LINEAR+CRTS baseline to $p=4.70\times10^{-3}$ after including ASAS-SN data (Table \ref{tab:drw}). The decrease in significance after adding new data is inconsistent with our expectation when a true periodic signal is present, which suggests that the periodic signal is not persistent.

The decrease in significance after including extended data was also seen for the sources in \cite{Charisi2016}. Their initial systematic search in the Palomar Transient Factory (PTF) identified $50$ periodic quasar candidates from $\sim 35,000$ spectroscopically-confirmed quasars.  They analyzed those candidates using additional data from CRTS and/or the intermediate Palomar Transient Factory (iPTF).  Of the $47$ candidates that have additional data, all but two had significantly increased p-values. Although the CRTS measurements have larger photometric uncertainties than PTF or iPTF and are in a different filter, the increase in the p-value may still be an indication that the additional data are inconsistent with the claimed periodicity. A similar phenomenon from the statistical perspective is also seen in a large sample of SDSS Stripe 82 quasars by \cite{Andrae2013}: although a small number of quasars are better described by the DRW+periodic model than the DRW-only model, more quasars are preferred by DRW-only as the number of observations increases. The failure of PG1302 and the many periodic candidates from \cite{Charisi2016} to demonstrate persistent periodicity therefore seems typical of the stochastic variability that is ubiquitous in normal (single black hole) quasars and AGN.

While quasar variability can be characterized by the DRW process, high frequency power law slopes that deviate from DRW have been found in a number of studies, including those using large samples from ground-based surveys \citep{Simm2016, Kozlowski2016, Caplar2017} and the ones using high quality \emph{Kepler} AGN light curves \citep{Mushotzky2011,Edelson2014,Aranzana2018,Smith2018}. Since assuming the incorrect PSD form would result in an overestimate of the significance of the periodic signal, we have also analyzed PG1302 under the more general, broken power law (BPL) model and take the PSD in Equation \ref{eqn:a} to be:

\begin{equation}
P(f) = \frac{Af^{-\alpha_{\rm lo}}}{1+(f/f_{\rm br})^{-\alpha_{\rm lo}+\alpha_{\rm hi}}}\quad,
\label{eqn:bpl}
\end{equation}

\noindent where $A$ is the normalization, $f_{\rm br}$ is the break frequency, and $\alpha_{\rm lo}$ and $\alpha_{\rm hi}$ are the low and high frequency slopes, respectively. The parameter ranges sampled are listed in Table \ref{tab:bpl}; as also shown in the table, while the BPL+periodic model is moderately preferred over the BPL only model and the best-fit period is consistent with that in the DRW+periodic model, evidence for the periodic signal also becomes weaker when ASAS-SN data are included.

\begin{deluxetable*}{lrrrr}
\tablecaption{Damped random walk parameter ranges sampled by MCMC \label{tab:drw}}
\tablehead{
\colhead{} & \colhead{DRW} & \colhead{DRW+periodic} & \colhead{DRW} & \colhead{DRW+periodic} \\
\colhead{} & \colhead{\tiny{(LINEAR+CRTS)}} & \colhead{\tiny{(LINEAR+CRTS)}} & \colhead{\tiny{(LINEAR+CRTS+ASASSN)}} & \colhead{\tiny{(LINEAR+CRTS+ASASSN)}}
}
\startdata
$\sigma_{\rm min}$ (mag day$^{-1/2}$)  & 0.00224 & 0.00224 & 0.00224 & 0.00224 \\
$\sigma_{\rm max}$ (mag day$^{-1/2}$)  & 0.0187 & 0.0187 & 0.0187 & 0.0187  \\
$\tau_{\rm min}$ (day)  & 200 & 200 & 200 & 200  \\
$\tau_{\rm max}$ (day)  &  3000 &  3000 & 3000 & 3000  \\
$\ln A_{\rm 0\,min}$  & \nodata & $-$10 & \nodata & $-$10  \\
$\ln A_{\rm 0\,max}$  & \nodata & 5 & \nodata & 5  \\
$P_{\rm min}$ (day)  & \nodata & 200 & \nodata & 200  \\
$P_{\rm max}$ (day)  & \nodata & 3000 & \nodata & 3000  \\
\hline
$\ln \mathcal{L}_{\rm max}$ & 20.49 & 27.07 & 33.19 & 38.55 \\
p-value & \nodata & 1.39$\times10^{-3}$ & \nodata & 4.70$\times10^{-3}$   \\
AIC & $-$36.98 & $-$46.15 &  $-$62.39 & $-$69.10 \\
P$_{\rm best fit}$ (day) & \nodata & $1773.60^{+434.87}_{-125.12}$ & \nodata & $2012.62^{+280.03}_{-219.96}$ \\
\enddata
\end{deluxetable*}

\begin{deluxetable*}{lrrrr}
\tablecaption{Broken power law parameter ranges sampled by MCMC \label{tab:bpl}}
\tablehead{
\colhead{} & \colhead{BPL} & \colhead{BPL+periodic} & \colhead{BPL} & \colhead{BPL+periodic} \\
\colhead{} & \colhead{\tiny{(LINEAR+CRTS)}} & \colhead{\tiny{(LINEAR+CRTS)}} & \colhead{\tiny{(LINEAR+CRTS+ASASSN)}} & \colhead{\tiny{(LINEAR+CRTS+ASASSN)}}
}
\startdata
A$_{\rm min}$ & 10$^{-3}$ & 10$^{-3}$ & 10$^{-3}$ & 10$^{-3}$ \\
A$_{\rm max}$ & 10$^{-2}$ & 10$^{-2}$ & 10$^{-2}$  & 10$^{-2}$ \\
$f_{\rm br\,min}$ (day$^{-1}$) & 0.00033 & 0.00033 & 0.00033 & 0.00033 \\
$f_{\rm br\,max}$ (day$^{-1}$) & 0.005 & 0.005 & 0.005 & 0.005 \\
$\alpha_{\rm lo\,min}$ & 0 & 0 & 0 & 0 \\
$\alpha_{\rm lo\,max}$ & 2 & 2 & 2 & 2 \\
$\alpha_{\rm hi\,min}$ & 2 & 2 & 2 & 2  \\
$\alpha_{\rm hi\,max}$ & 4 & 4 & 4 & 4 \\
$\ln A_{\rm 0\,min}$ & \nodata & $-$10 & \nodata & $-$10  \\
$\ln A_{\rm 0\,max}$ & \nodata & 5 & \nodata & 5 \\
$P_{\rm min}$ (day) & \nodata & 200 & \nodata & 200 \\
$P_{\rm max}$ (day) & \nodata & 3000 & \nodata & 3000  \\
\hline
$\ln \mathcal{L}_{\rm max}$ & 19.89 & 26.82 & 32.90 & 38.37  \\
p-value & \nodata & 9.78$\times10^{-4}$ & \nodata & 4.21$\times10^{-3}$ \\
AIC  & $-$31.79 & $-$41.65 & $-$57.80 & $-$64.75  \\
P$_{\rm best fit}$ (day) & \nodata & 1713.78$^{+133.90}_{-116.09}$ & \nodata & $1988.13^{+114.11}_{-845.88}$ \\
\enddata
\end{deluxetable*}


\section{Conclusions}\label{sec:conclude}

PG1302 has been reported as an SMBHB candidate, having shown apparent periodic variation over $\sim 2$ cycles on a LINEAR+CRTS baseline of $\sim 10$ years (G15). Its variability has been modeled as the relativistic Doppler boosting of the secondary mini-disk \citep{D'Orazio2015}, and it has an inferred binary separation of $\sim 0.01$ pc. If verified, PG1302 would be one of the most compact SMBHB candidates discovered yet, and searches using similar techniques can potentially uncover more candidates in the gravitational wave-emitting regime for multi-messenger studies with the pulsar timing arrays.

In this Letter, we have included the recent ASAS-SN data for this source, which has regular and dense sampling spanning $\sim 5$ years since CRTS and thus extends the total baseline to $\sim 15$ years. We have also applied a maximum likelihood analysis to search for a periodic component in addition to red noise, which is modeled as the DRW process or a BPL PSD. While we find that DRW+periodic or BPL+ periodic is the preferred model for the LINEAR+CRTS light curve, evidence for either model becomes weaker after adding ASAS-SN data. As Doppler boost from a binary should produce persistent periodicity, and more data should only strengthen the signal, our reanalysis suggests that the variability of PG1302 may be inconsistent with this proposed model. 

In this Letter, we have highlighted the importance of the long-term monitoring of SMBHB candidates that have been selected for their periodicity; it is also necessary to evaluate the significance of the periodic signal in the presence of stochastic variability. Any robust periodic quasar and SMBHB candidate should be able to withstand those two tests.


\acknowledgments

T.L. thanks A. Barth, M. Charisi, D. D'Orazio, Z. Haiman, D. Stern, and the anonymous referee for their comments. T.L. also thanks M. J. Graham for providing the archival data of PG1302. S.G. is supported in part by NSF AAG grant 1616566.

ASAS-SN is supported by the Gordon and Betty Moore Foundation through grant GBMF5490 to the Ohio State University and NSF grant AST-1515927. Development of ASAS-SN has been supported by NSF grant AST-0908816, the Mt. Cuba Astronomical Foundation, the Center for Cosmology and AstroParticle Physics at the Ohio State University, the Chinese Academy of Sciences South America Center for Astronomy (CAS-SACA), the Villum Foundation, and George Skestos.

The CSS survey is funded by the National Aeronautics and Space
Administration under Grant No. NNG05GF22G issued through the Science
Mission Directorate Near-Earth Objects Observations Program.  The CRTS
survey is supported by the U.S.~National Science Foundation under
grants AST-0909182.




\end{document}